\documentclass{vgtc}                          

\ifpdf
  \pdfoutput=1\relax                   
  \usepackage{graphicx}                
  \DeclareGraphicsExtensions{.pdf,.png,.jpg,.jpeg} 
\else
  \usepackage{graphicx}                
  \DeclareGraphicsExtensions{.eps}     
\fi%

\graphicspath{{figures/}{pictures/}{images/}{./}} 

\usepackage{microtype}                 
\PassOptionsToPackage{warn}{textcomp}  
\usepackage{textcomp}                  
\usepackage{mathptmx}                  
\usepackage{times}                     
\usepackage{cite}                      
\usepackage{tabu}                      
\usepackage{booktabs}                  

\onlineid{0}

\vgtccategory{Research}

\vgtcinsertpkg

\title{Aquanims -- Area-Preserving Animated Transitions\\ based on a Hydraulic Metaphor}

\author{Michael Aupetit\thanks{e-mail: maupetit@hbku.edu.qa}\\ %
        \scriptsize Qatar Computing Research Institute, HBKU, Doha, Qatar}

\abstract{We propose "\textit{Aquanims}" as new design metaphors for animated transitions that preserve displayed areas during the transformation. As liquids are incompressible fluids, we use a hydraulic metaphor to convey the sense of area preservation during animated transitions. We study the design space of Aquanims for rectangle-based charts.%
} 

\CCScatlist{ 
\CCScat{Computer Graphics}{I.3.3}{Picture/Image Generation}{Line and curve generation}
}

\begin{document}



\maketitle

\section{Introduction} 

Animated transitions are important in Information Visualization to keep the user linking and understanding seemingly different visualizations that transform into each other \cite{ChevalierRPCH16}. 

In statistical graphics, data exploration and understanding are supported by specific visual metaphors like bar charts, scatterplots, heatmaps or node-link diagrams. We focus on rectangle-based charts which use the size channel to encode the primary data. In these charts, the area of a graphical element is proportional to the underlying count, proportion or probability.


Usually, animated transitions between different area charts distort these areas: complete or partial occlusion modify the perceived area; standard linear interpolation between rectangles with different aspect ratio change their area; and some transformations are non trivial like changing the number of bins of an histogram while maintaining their total area through a continuous transition. 

In this work we consider that area is the key information supported by area charts and assume it is worth being preserved from a user perspective. We propose animated transitions that preserve these areas, and explore their design space for rectangle-based charts. In particular, our transitions obey an intuitive hydraulic metaphor, are not subject to occlusion and are applicable to any rectangle-based chart.

\section{Related work}

In \textit{SandDance} \cite{sanddance_Drucker2015} a metaphor based on particles is proposed. The animation consists in 3 stages: the source shape is segmented into particles representing unit data, the particles are moved, then they aggregate to form the target shape with same area. SandDance can be used to transition between many different charts including scatterplots and area-based charts. However the many particles moving is likely to disturb the perception of connection between specific source and target shapes. \textit{Visual Sedimentation} \cite{VisualSedimentation_Huron2013} is another design metaphor inspired by the physical process of sedimentation. It is especially suited to represent data stream. New data enter the visualization as bubbles attracted toward buckets, and progressively accumulate and create strata in form of color-coded area charts which summarize past data and progressively  compress the information stream. 

\section{Aquanims}

We propose a physical metaphor based on hydraulics to guide the design of area-preserving animated transitions. The rationale is twofold: first, liquids are incompressible fluids that are likely to convey the sense of volume preservation projected as areas on our retina; second, prospective human users are used to manipulate liquids in their everyday life: pouring milk in a glass, filling a sink, emptying a plastic bottle, transferring water between buckets or in a water tank, drinking a soda with a straw... not to mention experimental physics at school or in professional environments. We call \textit{\textit{Aquanims}} the animated transitions based on the proposed hydraulic metaphor.

Hydraulics is concerned with conveying liquids through pipes and channels, and to use it as a source of mechanical force or control. Typical hydraulic technology connects cylinders, tanks and pumps with pipes. As liquids are nearly incompressible, they keep their volume constant in the system, and they can transmit forces between distant containers connected by non-elastic pipes. Liquids have a mass so are subject to gravity forces, leading to equal surface level equilibrium in possibly distant but communicating vessels, as well as in natural artesian wells.

As tanks and vessels in hydraulic systems, area charts have containers like bars or tiles of which the area is measured. We will focus on bars and tiles of rectangular shape with axis-parallel edges in the sequel, although the analogy is not limited to these shapes and orientations and could be adapted in principle to animate pie charts for instance. Bars and tiles are usually the graphical objects whose area must be preserved as they encode the primary data. 

However in our physical analogy we consider that the equivalent to a bar or a tile is actually the liquid which takes the form of its graphical container. By so doing we transfer the semantic of the area of the graphical container to the one of the graphical fluid it contains. So we can change the shape of a set of containers, move the liquid from one container to another, or even let it be shared between several containers without changing its total volume. Thus we derive the two fundamental principles underlying \textit{Aquanims}: 
\textbf{First principle: the liquid encodes the data.} Data are mapped to the fluid  content rather than to its container. 
\textbf{Second principle: the liquid volume is constant}, so is its area in the graphical representation. Pipes and containers can change the liquid's shape but cannot change its total area. As a main guiding principle, \textit{Aquanims} should be designed so as to  emphasize the hydraulic metaphor evoking liquid-coded data (First principle), in order to increase the perception of area invariance (Second principle).
As the liquid rather than the container, encodes the data, it is color coded for data identification.

\section{\textit{Aquanims} building-blocks} 

\begin{figure*}
  \centering
  \includegraphics[width=\linewidth]{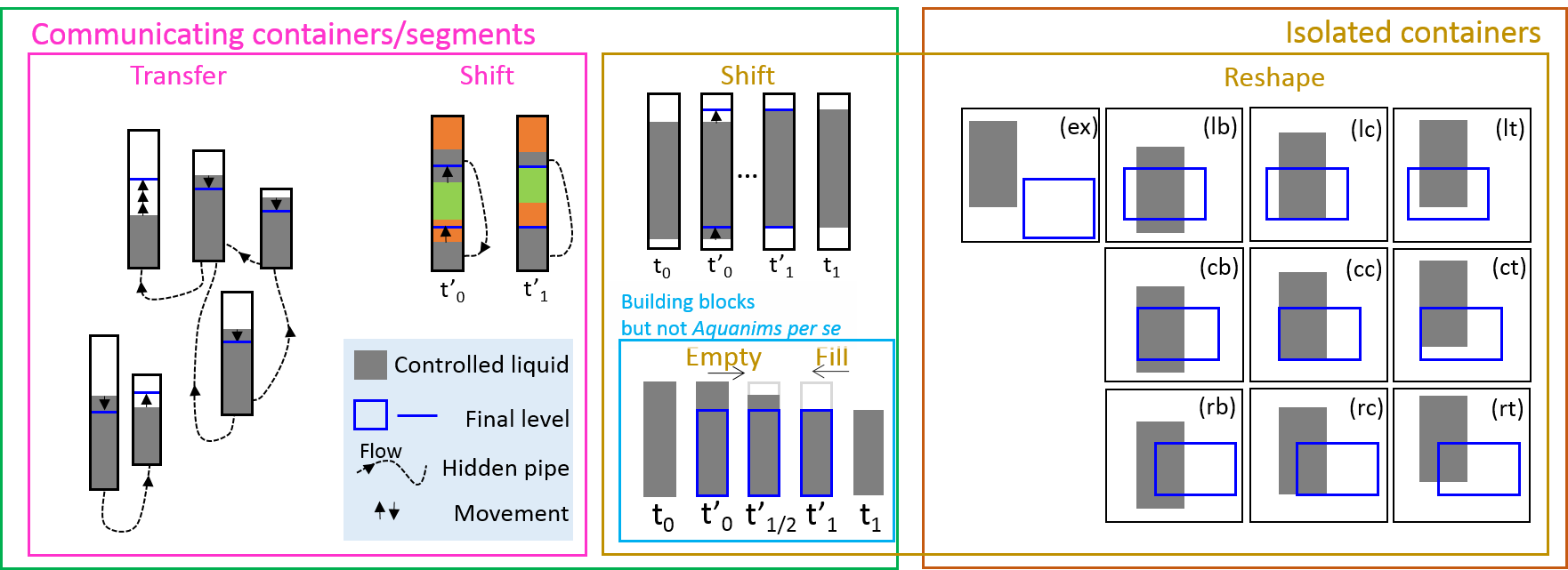}
  \caption{\label{fig:buildingblocks}%
           Building-blocks of \textit{Aquanims}. Filled gray and empty blue rectangles show the source and target states respectively. Both rectangles have the same area except for the building-blocks within the blue frame. The Reshape transitions involve 2 dimensions (2D) (Figure \ref{fig:reshapedetails}). }
\end{figure*}

We illustrate the different building-blocks that support the design of \textit{Aquanims} in the figure \ref{fig:buildingblocks}.

\begin{figure}
  \centering
  \mbox{} \hfill
  
  \includegraphics[width=\linewidth]{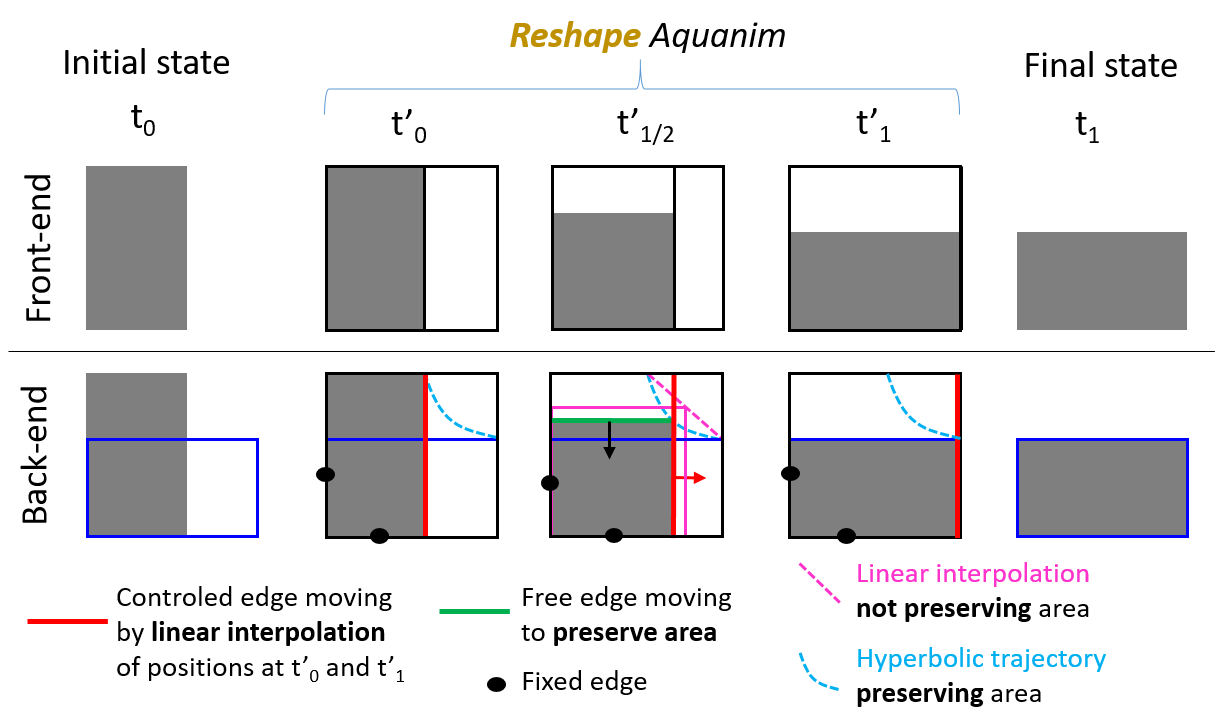}
  
  \caption{\label{fig:reshapedetails}%
          Reshape \textit{Aquanims} require non linear interpolations to keep constant areas. Here one side of the cylinder (red) moves horizontally using linear interpolation, the remaining free edge (green) moves vertically to maintain the constant area of the gray filled rectangle. }
\end{figure}

A single container can be \textbf{filled or emptied} and thus change area, but an Aquanim is formed when other containers are connected by invisible pipes to the former so the total area remains constant using \textbf{Transfer} of liquid between them. A \textbf{Shift} corresponds to a transfer of liquid from one segment to another in the same container.  
Regarding connected containers or segments, we have this formula between variation $h_k$ of levels within connected containers of respective width $w_k$: $\sum_k w_k h_k=0$. Assuming fixed-width containers like in bar charts and histograms, if the initial $L_k(0)$ and target $L_k(1)$ levels in each containers are such that total area $S=S(0)=\sum_k w_k L_k(0)=\sum_k w_k L_k(1)=S(1)$ is constant for these two states, then for any linearly interpolated state$L_k(t)=(1-u(t))L_k(0)+u(t)L_k(1)$, the area is preserved $S(t)=\sum_k w_k L_k(t)=S$.
Therefore standard linear interpolation  with possible slow-in/slow-out speed or any other speed scheme ensures the second principle of \textit{Aquanims}.

Isolated containers can also be \textbf{Reshaped} while preserving area using a hydraulic cylinder model (back-end) and aesthetic (front-end) as detailed in the figure \ref{fig:reshapedetails}. In that latter case, at least two orthogonal edges are moving during the transition.
In order to comply with the predictability principle of animated transitions \cite{HeerR07}, the start and end levels of the liquid in the container are remanent throughout the transition and materialize the hydraulic cylinder.   
Notice that linear interpolation between start and end positions of the rectangle vertices, does not preserve area (see figure \ref{fig:reshapedetails}). 
The general formula for the area-preserving \textbf{Reshape} animation is $w(t)=(1-u(t)) w(0) + u(t) w(1)$
and $ h(t)=\mathcal{A}/w(t)$ with $t\in[0,1]$, $w$ the width controlled directly by the slow-in/slow-out speed transition , $u(t)=3t^2-2t^3$ a cubic function generating the slow-in/slow-out speed for a regular sequence of increasing $t$ values. $h$ is the height of the rectangle indirectly controlled by the area-preserving condition. And $\mathcal{A}=w(0)h(0)=w(1)h(1)$ is the area of the rectangle, invariant during the \textit{Aquanim}.
 
\section{\textit{Aquanims} examples and future work}

\begin{figure}[htb]
\centering
\begin{tabular}{cccc}
\includegraphics[width=0.2\linewidth]{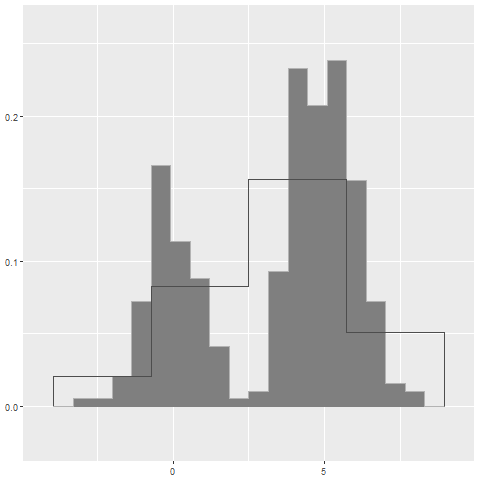}
&\includegraphics[width=0.2\linewidth]{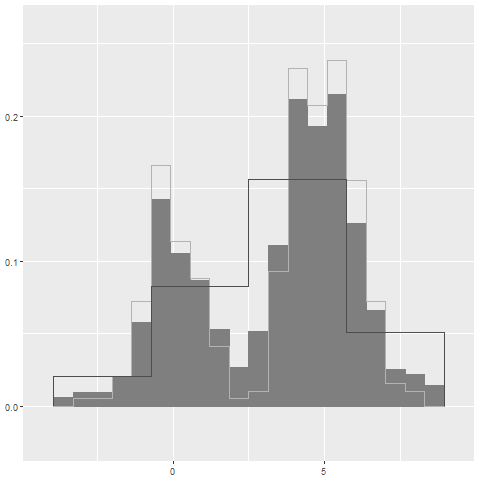}&
\includegraphics[width=0.2\linewidth]{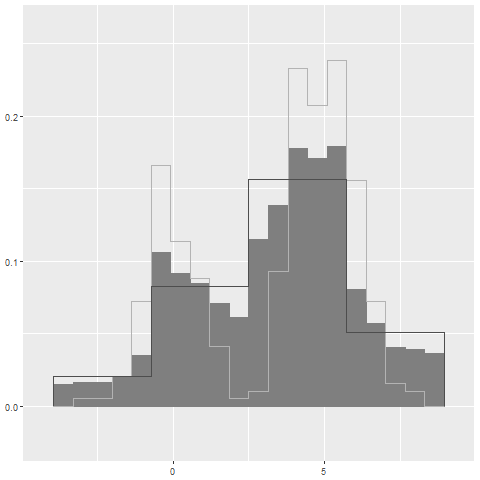}
&\includegraphics[width=0.2\linewidth]{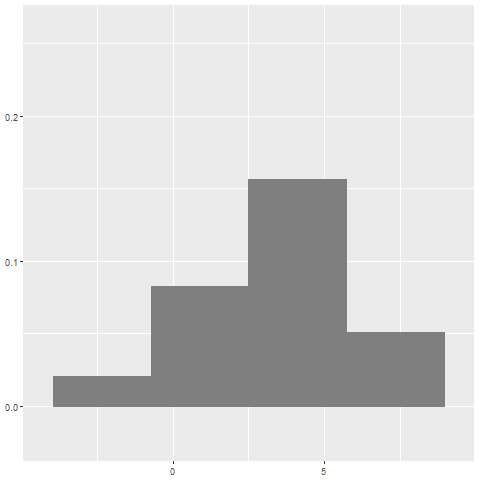}\\
\includegraphics[width=0.2\linewidth]{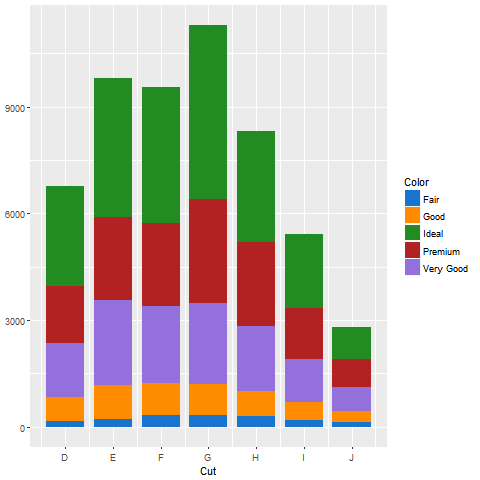}
&\includegraphics[width=0.2\linewidth]{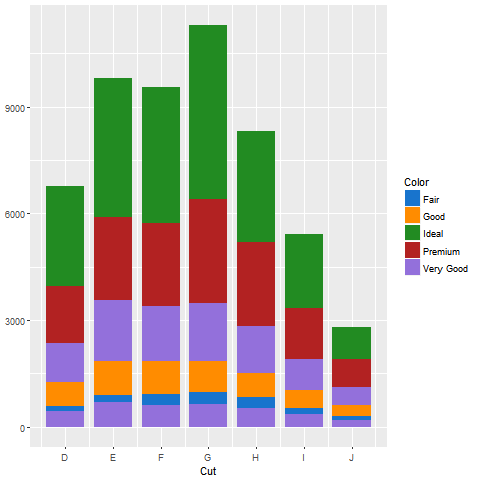}
&\includegraphics[width=0.2\linewidth]{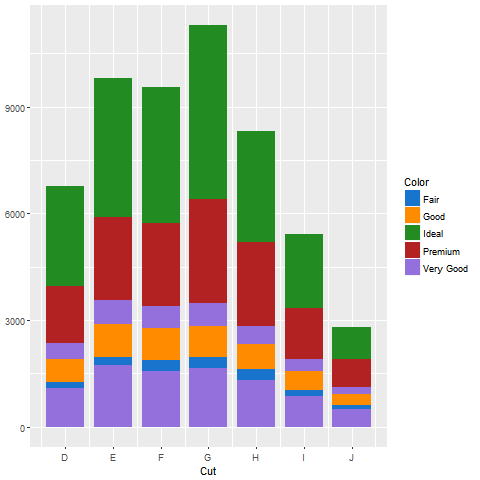}
&\includegraphics[width=0.2\linewidth]{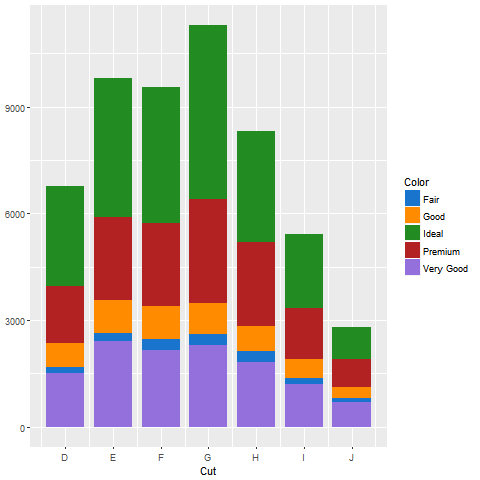}\\
\end{tabular}
\caption{\label{fig:adapchangingnumbins}%
\textit{Aquanim} for changing number of histogram bins (top) and for aligning selected segments (magenta) on the x-axis baseline (bottom). }
\vspace{-.5cm}
\end{figure}

We develop the animations using R and Shiny softwares and the \textit{ggplot2} and \textit{animation} packages for the graphics. 
The \textbf{Transfer} building-block is used in the figure \ref{fig:adapchangingnumbins}  (top) to animate a change in the number of bins of a histogram, a typical task when attempting to visually approximate the data probability density function . The \textbf{Shift} building-block is used in the figure \ref{fig:adapchangingnumbins} (bottom) to animate the alignment of some segments to the x-axis baseline in a stacked bar chart, a typical task when their relative height must be compared. 
Other examples are given in supplementary material. 

How to adapt Aquanims to non rectangle-based charts and how effective are these animations for the user, are still open questions.

\bibliographystyle{abbrv-doi}

\bibliography{egbibsample}
\end{document}